# Current status of space gravitational wave antenna DECIGO and B-DECIGO

Seiji Kawamura[1,*], Masaki Ando[2], Naoki Seto[3], Shuichi Sato[4], Mitsuru Musha[5], Isao Kawano[6], Jun'ichi Yokoyama[7], Takahiro Tanaka[3], Kunihito Ioka[8], Tomotada Akutsu[9], Takeshi Takashima[10], Kazuhiro Agatsuma[11], Akito Araya[12], Naoki Aritomi[2], Hideki Asada[13], Takeshi Chiba[14], Satoshi Eguchi[15], Motohiro Enoki[16], Masa-Katsu Fujimoto[9], Ryuichi Fujita[17], Toshifumi Futamase[18], Tomohiro Harada[19], Kazuhiro Hayama[15], Yoshiaki Himemoto[20], Takashi Hiramatsu[19], Feng-Lei Hong[21], Mizuhiko Hosokawa[22], Kiyotomo Ichiki[1], Satoshi Ikari[23], Hideki Ishihara[24], Tomohiro Ishikawa[1], Yousuke Itoh[24], Takahiro Ito[25], Shoki Iwaguchi[1], Kiwamu Izumi[10], Nobuyuki Kanda[24], Shinya Kanemura[26], Fumiko Kawazoe[27], Shiho Kobayashi[28], Kazunori Kohri[29], Yasufumi Kojima[30], Keiko Kokeyama[31], Kei Kotake[15], Sachiko Kuroyanagi[1], Kei-ichi Maeda[32], Shuhei Matsushita[23], Yuta Michimura[2], Taigen Morimoto[1], Shinji Mukohyama[8], Koji Nagano[10], Shigeo Nagano[22], Takeo Naito[1], Kouji Nakamura[9], Takashi Nakamura[3], Hiroyuki Nakano[33], Kenichi Nakao[24], Shinichi Nakasuka[23], Yoshinori Nakayama[34], Kazuhiro Nakazawa[35], Atsushi Nishizawa[7], Masashi Ohkawa[36], Kenichi Oohara[37], Norichika Sago[3], Motoyuki Saijo[38], Masaaki Sakagami[39], Shin-ichiro Sakai[10], Takashi Sato[40], Masaru Shibata[8,41], Hisaaki Shinkai[42], Ayaka Shoda[9], Kentaro Somiya[43], Hajime Sotani[44], Ryutaro Takahashi[9], Hirotaka Takahashi[45], Takamori Akiteru[12], Keisuke Taniguchi[46], Atsushi Taruya[8], Kimio Tsubono[2], Shinji Tsujikawa[32], Akitoshi Ueda[47], Ken-ichi Ueda[5], Izumi Watanabe[1], Kent Yagi[48], Rika Yamada[1], Shuichiro Yokoyama[35], Chul-Moon Yoo[1], Zong-Hong ZHU[49]

*[1] Division of Particle and Astrophysical Sciences, Nagoya University, Nagoya, Aichi, 464-8602, Japan*
*[2] Department of Physics, The University of Tokyo, Bunkyo, Tokyo 113-0033, Japan*
*[3] Department of Physics, Graduate School of Science, Kyoto University, Kyoto, Kyoto 606-8502, Japan*
*[4] Faculty of Science and Engineering, Hosei University, Koganei, Tokyo 184-8584, Japan*
*[5] Institute for Laser Science, The University of Electro-Communications, Chofu, Tokyo 182-8585, Japan*
*[6] Japan Aerospace Exploration Agency*
*Tsukuba Space Center, Tsukuba, Ibaraki 305-8505, Japan*
*[7] Research Center for the Early Universe, School of Science, The University of Tokyo, Bunkyo, Tokyo 113-0033, Japan*
*[8] Yukawa Institute for Theoretical Physics, Kyoto University, Kyoto, Kyoto 606-8502, Japan*
*[9] Gravitational Wave Science Project, National Astronomical Observatory of Japan, Mitaka, Tokyo 181-8588, Japan*
*[10] Institute of Space and Astronautical Science, Japan Aerospace Exploration Agency, Sagamihara, Kanagawa 252-5210, Japan*
*[11] School of Physics and Astronomy and Institute for Gravitational Wave Astronomy,*
*University of Birmingham, Edgbaston, Birmingham B15 2TT, UK*
*[12] Earthquake Research Institute, The University of Tokyo, Bunkyo, Tokyo 113-0032, Japan*
*[13] Faculty of Science and Technology, Hirosaki University, Hirosaki , Aomori 036-8561, Japan*
*[14] Department of Physics, College of Humanities and Sciences, Nihon University, Setagaya, Tokyo 156-8550, Japan*
*[15] Department of Applied Physics, Faculty of Science, Fukuoka University, Fukuoka, Fukuoka 814-0180, Japan*
*[16] Faculty of Business Administration, Tokyo Keizai University, Kokubunji, Tokyo 185-8502, Japan*
*[17] Institute of Liberal Arts, Otemon Gakuin University, Ibaraki, Osaka 567-8502, Japan*
*[18] Kyoto Sangyo University, Kyoto, Kyoto 603-8555, Japan*
*[19] Department of Physics, Rikkyo University, Toshima, Tokyo 171-8501, Japan*
*[20] College of Industrial Technology, Nihon University, Narashino, Chiba 275-8576, Japan*
*[21] Department of Physics, Yokohama National University, Yokohama, Kanagawa 240-8501, Japan*

[22] National Institute of Information and Communications Technology (NICT), Koganei, Tokyo 184-8795, Japan
[23] Department of Aeronautics and Astronautics, The University of Tokyo, Hongo, Bunkyo, Tokyo 113-8656 Japan
[24] Department of Physics, Osaka City University, Osaka, Osaka 558-8585, Japan
[25] Japan Aerospace Exploration Agency, Sagamihara, Kanagawa 252-5210, Japan
[26] Department of Physics, Osaka City University, Osaka, Osaka 558-8585, Japan
[27] Institut fur Gravitationsphysik Leibniz Universitat Hannover, Callinstr. 38, 30167 Hannover, Germany
[28] Astrophysics Research Institute, 146 Brownlow Hill, Liverpool L3 5RF, UK
[29] Institute of Particle and Nuclear Studies, KEK, Tsukuba, Ibaraki 305-0801, Japan
[30] Graduate School of Science, Hiroshima University, Higashi-hiroshima, Hiroshima 739-8526, Japan
[31] KAGRA Observatory, ICRR, The University of Tokyo, Hida, Gifu 506-1205, Japan
[32] Department of Physics, Graduate School of Advanced Science and Engineering, Waseda University, Shinjuku, Tokyo, 169-8555, Japan
[33] Faculty of Law, Ryukoku University, Kyoto, Kyoto 612-8577, Japan
[34] Department of Aerospace Engineering, National Defense Academy, Yokosuka, Kanagawa 239-8686, Japan
[35] The Kobayashi-Maskawa Institute for the Origin of Particles and the Universe, Nagoya University, Nagoya, Aichi 464-8602, Japan
[36] Faculty of Engineering, Niigata University, Niigata, Niigata 950-2181, Japan
[37] Department of Physics, Niigata University, Niigata, Niigata 950-2181, Japan
[38] Research Institute for Science and Engineering, Waseda University
Shinjuku, Tokyo 169-8555, Japan
[39] Graduate School of Human and Environmental Studies, Kyoto University, Kyoto, Kyoto 606-8501, Japan
[40] Niigata College of Technology, Niigata, Niigata 950-2076, Japan
[41] Max Planck Institute for Gravitational Physics at Potsdam, Am Mu˙hlenberg 1, D-14476 Potsdam-Golm, Germany
[42] Dept of Information Systems, Osaka Institute of Technology, Kitayama, Hirakata 573-0196, Japan
[43] Department of Physics, Tokyo Institute of Technology, Meguro, Tokyo 152-8551, Japan
[44] RIKEN, Wako, Saitama 351-0198, Japan
[45] Department of Management and Information Systems Engineering, Nagaoka University of Technology, Nagaoka, Niigata 940-2188, Japan
[46] Department of Physics, University of the Ryukyus, Nishihara, Okinawa 903-0213, Japan
[47] JASMINE Project, National Astronomical Observatory of Japan, Mitaka, Tokyo 181-8588, Japan
[48] University of Virginia, Department of Physics, Charlottesville, VA 22904, USA
[49] Beijing Normal University, Haidian District, Beijing 100875, China

*E-mail: kawamura@u.phys.nagoya-u.ac.jp

## Abstract

Deci-hertz Interferometer Gravitational Wave Observatory (DECIGO) is the future Japanese space mission with a frequency band of 0.1 Hz to 10 Hz. DECIGO aims at the detection of primordial gravitational waves, which could be produced during the inflationary period right after the birth of the universe. There are many other scientific objectives of DECIGO, including the direct measurement of the acceleration of the expansion of the universe, and reliable and accurate predictions of the timing and locations of neutron star/black hole binary coalescences. DECIGO consists of four clusters of observatories placed in the heliocentric orbit. Each cluster consists of three spacecraft, which form three Fabry-Perot Michelson interferometers with an arm length of 1,000 km. Three clusters of DECIGO will be placed far from each other, and the fourth cluster will be placed in the same position as one of the three clusters to obtain the correlation signals for the detection of the primordial gravitational waves. We plan to launch B-DECIGO, which is a scientific pathfinder of DECIGO, before DECIGO in the 2030s to demonstrate the technologies required for DECIGO, as well as to obtain fruitful scientific results to further expand the multi-messenger astronomy.

Subject Index [Insert subject index codes here]

## 1. Introduction

Gravitational waves were detected for the first time by the Laser Interferometer Gravitational-wave Observatory (LIGO) in 2015 [1]. They found that the gravitational waves came from the merger of the black hole binary that occurred ~400 Mpc away from the earth. This detection successfully established gravitational wave astronomy. After several subsequent detections of gravitational waves from the black hole binary inspirals, in 2017, LIGO, together with Virgo, caught gravitational waves emitted from the merger of a neutron star binary [2]. This detection led to successful follow-up observations with electromagnetic waves [3] [4]. This event not only revealed the mechanism of the r-process that occurred around the neutron star binary coalescence but also opened multi-messenger astronomy. Since then, LIGO and Virgo have improved the sensitivity of the detectors significantly; during the latest observation run, the detection became routine, once or twice a week. Soon the Large-scale Cryogenic Gravitational Wave Telescope (KAGRA) [5] will join LIGO and Virgo to improve the quality of the gravitational wave detection. Furthermore, LIGO India will eventually join the world network.

There are also more challenging plans in the future: Einstein Telescope (ET) [6] and Cosmic Explorer (CE) with a longer arm length of 10 km in a triangular arrangement, and 40 km as an ‘L’, respectively. They plan to accomplish significantly better sensitivity than the current ground-based detectors at frequencies down to a few Hz. In space, the European Space Agency (ESA) has a future mission called Laser Interferometer Space Antenna (LISA) [7] in collaboration with the National Aeronautics and Space Administration (NASA). LISA consists of three spacecraft with an arm length of 2.5 million km. LISA aims at detecting gravitational waves coming from intermediate-mass to massive black hole binaries at frequencies between 1 mHz to 0.1 Hz. ESA and NASA launched the LISA Pathfinder [8] in 2015 and attained the acceleration noise requirement for LISA. Encouraged by these successful results, they plan to launch LISA in 2034.

In Japan, we have also been considering space missions: the Deci-hertz Interferometer Gravitational Wave Observatory (DECIGO) [9] [10], and its scientific pathfinder B-DECIGO [11]. The Japanese Gravitational Wave Community (JGWC) already agreed in 2018 that we regard observations of gravitational waves by KAGRA as the highest priority, and after that, we develop gravitational wave astronomy by DECIGO and B-DECIGO. In this paper, we will explain the design, objectives, and schedule of DECIGO and B-DECIGO in detail.

## 2. Design of DECIGO and B-DECIGO

### *2.1. General concept*

DECIGO consists of four clusters of observatories. As shown in Fig. 1, we plan to put them into the heliocentric orbit in such a way that two clusters are placed in the same position, and the other two clusters are distributed around the sun. As shown in Fig. 2, each cluster consists of three spacecraft, which form an equilateral triangle with a side of 1,000 km. Each spacecraft, which has a drag-free system, contains two mirrors floating inside the satellite as a proof masses. DECIGO measures a change in the distance caused by gravitational waves between the two mirrors, which are 1,000 km apart from each other. We employ a Fabry-Perot (FP) cavity, consisting of the two mirrors, to measure the arm length change. The target frequency band of DECIGO is between 0.1 Hz and 10 Hz. The mission lifetime of DECIGO is at least three years.

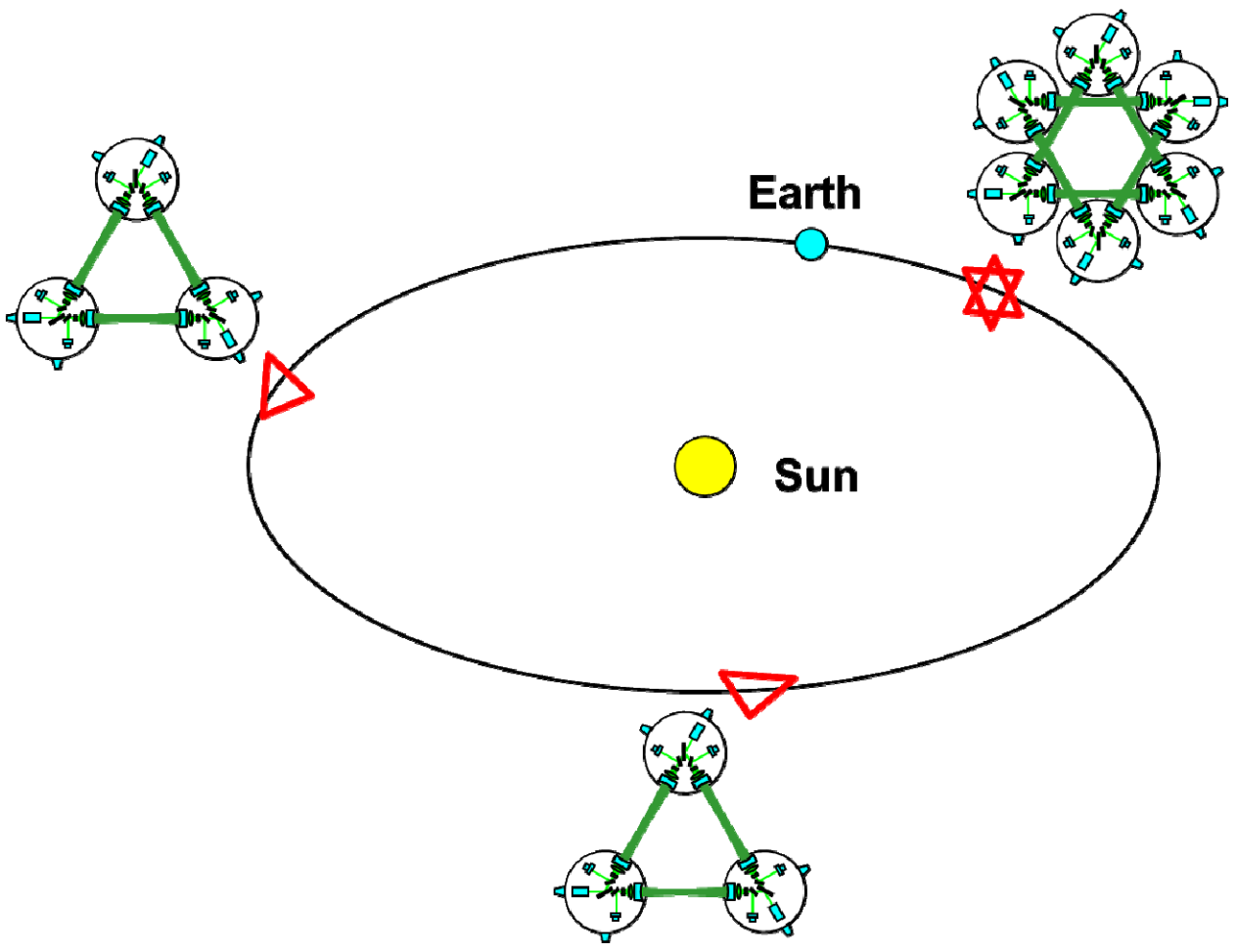

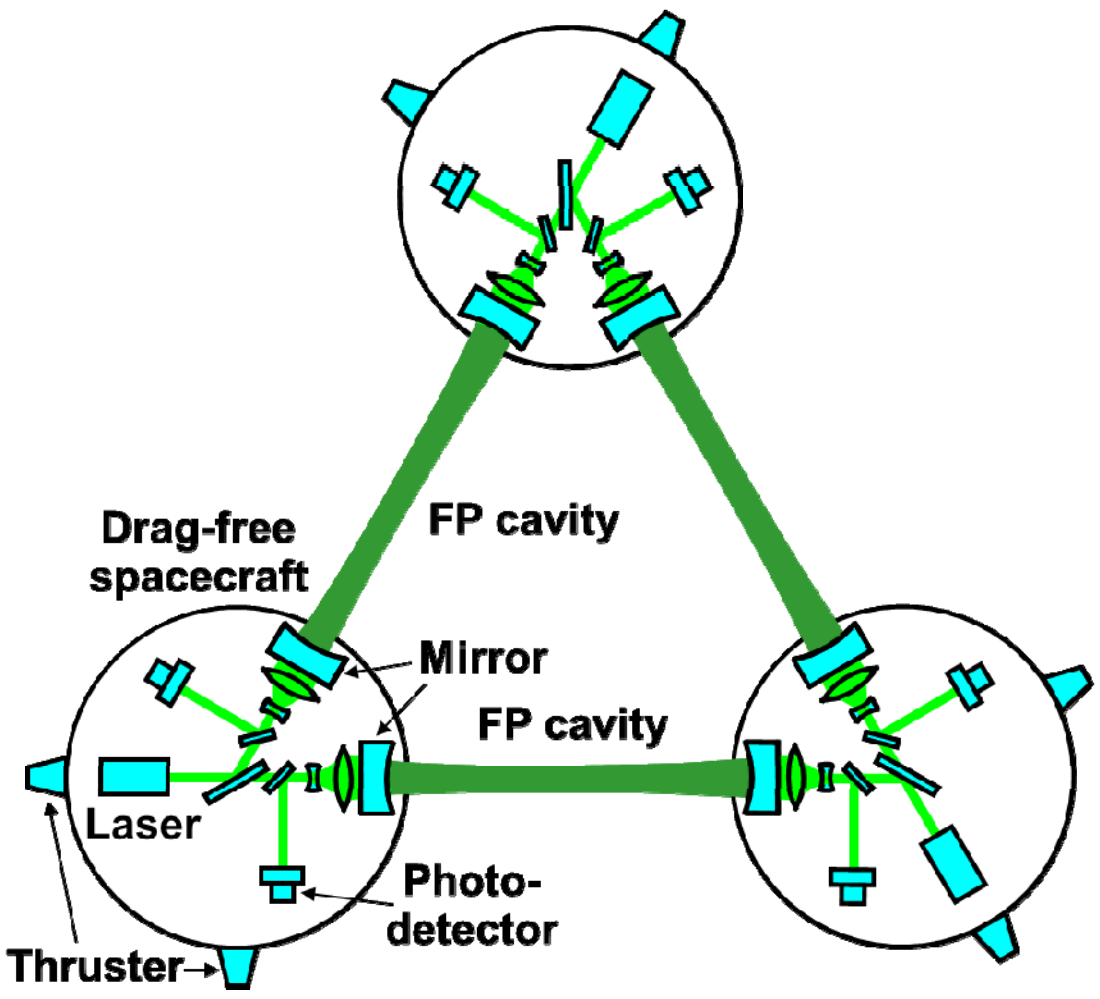

Fig. 1. Orbit of DECIGO. Four clusters of DECIGO are put in the heliocentric orbit: two at the same position and the other two at different positions.

Fig. 2. Conceptual design of DECIGO. One cluster of DECIGO consists of three drag-free spacecraft. FP cavities are used to measure a change in the arm length.

The reason for the above design of DECIGO is the following. First, let us explain the bandwidth of DECIGO. We chose a frequency band of DECIGO to be 0.1 Hz to 10 Hz because this frequency band is located in a gap between the frequency band of LISA and the ground-based detectors. It means that DECIGO can be used as a follow-up detector after LISA detects gravitational waves coming from the intermediate-mass black hole binaries. Also, DECIGO can be used as a forecast detector before the ground-based detectors detect gravitational waves coming from the 10-100 solar-mass black hole binaries or neutron star binaries.

Another reason that justifies this frequency band comes from the consideration of the importance of low-frequency gravitational waves and the confusion limiting noise. Generally speaking, we can say that at a lower frequency regime, the expected number of gravitational-wave sources is larger, relative to the available data amount. This is because, at lower frequencies, the evolution timescale becomes longer, and we also have a larger variety of gravitational wave sources. Up to ~0.1Hz, the confusion limiting noise is dominated by a large number of gravitational wave signals coming from white dwarf binaries. These gravitational wave signals are so abundant that many more than two signals are likely to exist within one frequency resolution bin, which is determined by the observation time. As a result, separation of these signals becomes a significant challenge, and they form a sensitivity limit for signals in that frequency range. This confusion limiting noise due to white dwarf binaries exists only below ~0.1 Hz because they merge when their gravitational waves reach ~0.1 Hz. While we still need to deal with gravitational waves from binaries composed of black holes and neutron stars, it is expected that, above ~0.1 Hz, we can open a deep window without this confusion limit of gravitational waves.

It is also essential to point out that the expected strain of the primordial gravitational waves is relatively reachable around 0.1 Hz, compared with a higher frequency regime. Therefore, the frequency band between 0.1 Hz and 10 Hz is the most fruitful band for future gravitational wave astronomy and physics.

### *2.2. Design of DECIGO*

Now we optimize the arm length for the frequency band. We chose the arm length of DECIGO to be 1,000 km. This size is much longer than ground-based detectors, but it is much shorter than LISA. We can justify this choice by the consideration of optimizing the quantum noise, which consists of shot noise and radiation pressure noise, for this frequency band. If the arm length is long, the laser light interacts with the gravitational waves for a longer time, resulting in larger arm length changes due to gravitational wave signals. On the other hand, only a fraction of the light can reach the other mirror because of the diffraction of light, resulting in larger relative fluctuations of the shot noise. These two effects on signals and shot noise cancel each other [10]. Therefore, the shot noise in terms of strain (simply called "shot noise" hereafter) does not depend on the arm length.

Incidentally, the shot noise is white in frequencies below a frequency determined by the propagation time of the light in the arms. This is because a part of the gravitational wave signals cancels above the corner frequency. More precisely, the corner frequency of the shot noise is inversely proportional to the arm length of the detector.

As for the radiation pressure noise, shorter arm length increases the radiation pressure noise in terms of strain (simply called "radiation pressure noise" hereafter) because of the two reasons: higher illuminating laser power and lower signals. As a result, the crossover frequency between the radiation pressure noise and shot noise is inversely proportional to the arm length of the detector.

Altogether, the quantum noise in terms of strain (simply called "quantum noise" hereafter) shifts to lower frequencies with a shorter arm length, maintaining the noise curve shape in the logarithmic scale. Moreover, if the arm length is short enough, we can even implement FP cavities to enhance the gravitational wave signals, thus to improve the shot noise at the expense of the radiation pressure noise and the bandwidth. With the above considerations, we found that an interferometer with an arm length of 1,000 km, a mirror diameter of 1 m, and a laser wavelength of 515 nm accommodates a FP cavity with a finesse of 10 (See Table 1).

Table 1. Optical and mechanical parameters of DECIGO and B-DECIGO [10].

| Optical/mechanical parameter | DECIGO | B-DECIGO |
|---|---|---|
| Arm length (km) | 1,000 | 100 |
| Laser power (W) | 10 | 1 |
| Laser wavelength (nm) | 515 | 515 |
| Finesse | 10 | 100 |
| Mirror diameter (m) | 1 | 0.3 |
| Mirror mass (kg) | 100 | 30 |
| # of clusters | 4 | 1 |
| # of interferometers per cluster | 3 | 3 |

Other parameters of DECIGO are as follows. We chose the mass of the mirror to be 100 kg, and the laser power 10 W to optimize the quantum noise around 0.1 Hz (See Table 1). As a result, the radiation pressure noise is larger than the shot noise below 0.1 Hz, and the shot noise is white between 0.1 Hz and 10 Hz, and above 10 Hz the shot noise increases linearly with the frequency.

Another crucial element of DECIGO technologies is a drag-free system. A spacecraft receives non-gravitational forces from various entities, such as solar radiation and cosmic dust. As a result, the motion of the spacecraft due to these "drag" would be much larger than the motion change caused by gravitational waves. Thus, we need to employ a drag-free system. The drag-free system requires a proof mass floating inside the spacecraft. A local sensor implemented inside the spacecraft measures the relative position of the proof mass with respect to the spacecraft. Then the relative position signal is fed back to the thrusters attached to the spacecraft. The drag-free system controls the position of the spacecraft to the proof mass. In the case of DECIGO, the mirrors play a role as proof masses. The drag-free system requires accurate control of the spacecraft to the mirrors because a change in the gravitational field by the motion of the spacecraft could shake the mirrors.

Now let us explain a bit more details about the interferometer and cluster configuration of DECIGO. As shown in Fig. 2, one cluster of DECIGO, which consists of three spacecraft, has three equilateral-triangular FP Michelson interferometers. Each FP Michelson interferometer has two FP cavities. Each FP cavity is shared by two different FP Michelson interferometers. Implementing two independent interferometers in one cluster is crucial to obtain the gravitational wave signals with different polarizations. The third interferometer is redundant; it plays a role as a backup in case one of the other two interferometers fails. We could also look at the null signal by combining the three interferometer outputs to improve the confidence in the detected gravitational wave signals.

We plan to place four clusters of DECIGO in the heliocentric orbit (See Fig. 1). We will place three clusters equally spaced around the sun to ensure the best angular resolution of the gravitational wave sources. We will place the fourth cluster together with one of the three clusters to obtain the correlation signals for the detection of the primordial gravitational waves. Since the noise of each DECIGO cluster is independent, while the gravitational wave signals are common to two different clusters, correlating two cluster outputs improves the effective sensitivity of DECIGO. We plan to take the correlation signal for at least three years.

We plan to accomplish the quantum-noise-limited sensitivity for DECIGO by reducing all the other noises, such as the thermal noise of the mirrors and the gravitational field fluctuations. As shown in Fig. 3, the target strain sensitivity of one cluster of DECIGO is about $4 \times 10^{-24}$ $Hz^{-1/2}$ around 1 Hz, and the target three-year-correlation strain sensitivity is about $7 \times 10^{-26}$ $Hz^{-1/2}$ around 1 Hz.

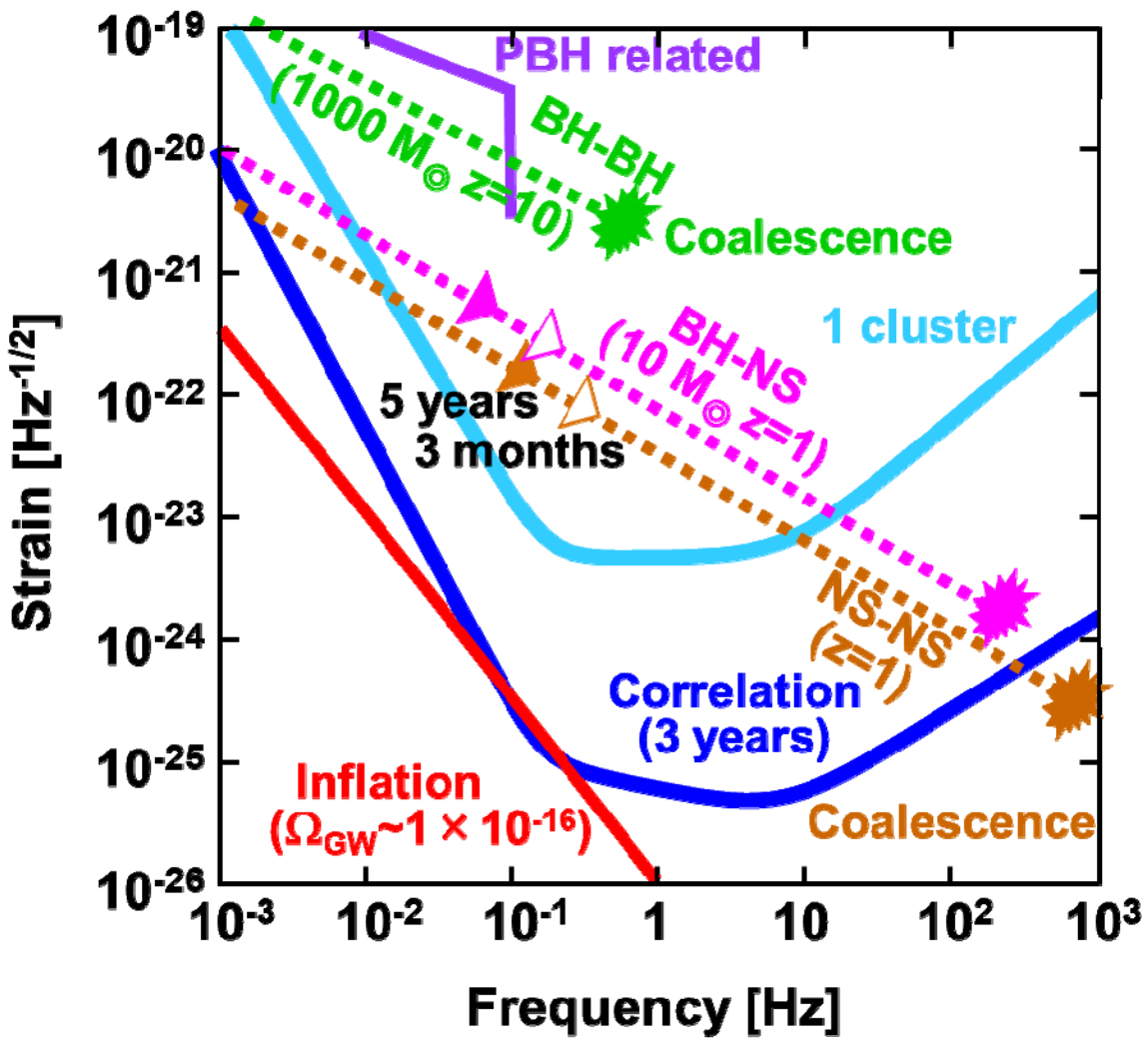

Fig. 3. Target strain sensitivity of one cluster of DECIGO (light blue line) and correlation of two clusters for 3 years (dark blue line), together with expected gravitational-wave strain signals: primordial gravitational waves corresponding to $\Omega_{GW}$~$1 \times 10^{-16}$ (red line), black hole (1,000 solar mass) binary at z=10 (BH-BH; green dotted line), black hole (10 solar mass) – neutron star binary at z=1 (BH-NS; pink dotted line; filled and open triangles indicate 5 years and 3 months before the coalescence, respectively.), neutron star binary at z=1 (NS-NS; brown dotted line; filled and open triangles indicate 5 years and 3 months before the coalescence, respectively.), and primordial dark matter-related gravitational waves (PBH; purple line).

### *2.3. Design of B-DECIGO*

We plan to launch B-DECIGO before DECIGO to expand the gravitational wave astronomy and physics, as well as to demonstrate the technologies required to realize DECIGO. As shown in Fig. 1, B-DECIGO has one cluster, which consists of three drag-free spacecraft. The cluster has three FP Michelson interferometers in the same manner as DECIGO. The arm length of B-DECIGO is 100 km. The finesse of the FP cavities is 100, with the mirror diameter of 0.3 m and the laser wavelength of 515 nm. We chose interferometer parameters with a laser power of 1 W and a mirror mass of 30 kg. Overall, the technologies required for B-DECIGO are slightly less challenging than DECIGO. The orbit of B-DECIGO is still to be determined. The target strain sensitivity of B-DECIGO is $6 \times 10^{-23}$ $Hz^{-1/2}$ around 1 Hz.

## 3. Objectives of DECIGO and B-DECIGO

### *3.1. Objectives of DECIGO*

There are many science targets that DECIGO aims at, as shown in Fig. 3. The most important scientific objective of DECIGO is to detect the primordial gravitational waves, which could be produced during the inflation period [12] [13] [14] right after the birth of the universe. In the standard inflation theory, stochastic gravitational waves are generated as quantum fluctuations of space-time. In the commoving frame, the wavelength of a gravitational wave mode does not change with time, but the Hubble horizon shrinks during the inflation phase. The quantum nature of a gravitational wave mode was lost, when the horizon size decreased to a size comparable to its wavelength. After the horizon crossing, the mode was frozen (not in the sense of cryogenic freezing). At the end of inflation, the Hubble horizon began to expand in the commoving frame. When the horizon size again became larger than the wavelength, the gravitational wave mode was defrosted and started oscillating and propagating as a standard gravitational wave. Finally, these primordial gravitational waves reach us now.

The primordial gravitational waves with a larger wavelength extended beyond the horizon size at the earlier stage of the inflation period, and the wavelength relatively diminished below the horizon size at the more recent time. On the other hand, the primordial gravitational waves with a shorter wavelength extended beyond the horizon size at the later stage of the inflation period, and the wavelength relatively diminished below the horizon size at an ealier time. The frequency of the primordial gravitational waves, which we aim at detecting with DECIGO, is around 0.1 Hz. Thus we could directly observe the inflation at a particular time, corresponding to the frequency of 0.1 Hz. Figure 3 indicates that DECIGO could barely detect the expected primordial gravitational waves from this inflationary scenario with the three-year correlation method. We plan to optimize the optical and mechanical parameters of DECIGO to improve the sensitivity further to ensure the solid detection of the primordial gravitational waves or a significant upper limit.

There are many variants of the inflation models, and there is even a possibility that the inflation did not take place. DECIGO could capture strong evidence supporting inflation and also provide us with crucial observational measures for discriminating inflation models. For example, DECIGO could determine the energy scale of the inflation as well as the reheating temperature of the inflation, depending on model parameters. Here we should emphasize that electromagnetic waves started propagating straight after 380,000 years after the birth of the universe. Thus, it is impossible to observe the universe beyond that time directly with light; the primordial gravitational waves are highly penetrating and can be regarded as a precious fossil with which we can uniquely and directly access the beginning of the universe.

Another important objective of DECIGO is to measure the acceleration of the expansion of the universe without relying on astronomical empirical rules. The principle of the measurement is straightforward [9]. We just measure the phase and amplitude of the waveform of the gravitational waves coming from the neutron star binary at a far distance from the earth (See Fig. 3). We can determine the distance to the binary by comparing the observed amplitude and the phase. In addition, the waveform would show a phase change caused by the acceleration of the expansion of the universe. These measurements will be especially valuable for the investigation of dark energy because they are based only on fundamental physics laws.

As for the detection of gravitational waves coming from the neutron star binaries, DECIGO will provide a reliable and accurate prediction for their coalescences. DECIGO would catch ~100,000 gravitational wave events per year coming from the neutron star binaries within a redshift of 5. The observable signals are up to several years in advance before the coalescences. The angular accuracy of the determination of the direction of the cosmological gravitational wave sources could be as small as ~1 arcsec (based on [15]). The accuracy of the predicted coalescence time could be typically ~0.1 sec. This accurate prediction enables simultaneous gamma-ray observations and electromagnetic follow-ups to be more reliable and frequent. We can expect that the multi-messenger astronomy will be developed significantly because of DECIGO.

The objectives of DECIGO also include the following (See Fig. 3). DECIGO could reveal the mechanism of the formation of massive black holes in the center of galaxies, by detecting abundant gravitational waves coming from the intermediate-mass black hole binaries. DECIGO is useful to examine the accuracy of general relativity [16]. For example, by detecting gravitational waves coming from neutron star and 10-solar-mass black hole binaries at a redshift of 1, the constraint on the Brans-Dicke parameter would be improved by four orders of magnitude compared with the current limit obtained by the measurement of Cassini. DECIGO could separate the scalar and vector modes, which could have been produced at the beginning of the universe, from the tensor mode of gravitational waves [17]. DECIGO could examine the parity symmetry between the clockwise and counter-clockwise gravitational-wave polarizations by placing the two clusters of DECIGO with a slight distance [18]. DECIGO could determine whether the primordial black holes are a contributor to dark matter by detecting gravitational waves emitted from the fluctuations of the mass density at the beginning of the universe [19]. DECIGO could detect gravitational waves from a black hole binary merging in the ground-based detector's frequency band with precise parameter estimations, especially on the coalescing time and sky location [20]. DECIGO could provide a meaningful upper limit for the number and mass of the scalar field by detecting gravitational waves emitted from the electroweak phase transition [21].

### *3.2. Objectives of B-DECIGO*

The objectives of B-DECIGO are mainly astrophysical ones. B-DECIGO will predict the time and direction of the neutron star binary coalescences ~100 times a year. The angular accuracy of the direction of the sources is about ~0.01 $deg^2$ (scaling results in [22] using [23]). The timing accuracy of the coalescences is about ~0.1 sec for the sources at a redshift of 0.1. This accurate prediction provides useful information to contribute significantly to multi-messenger astronomy. Moreover, B-DECIGO could reveal the formation mechanism of 30 solar mass black holes, which were detected by LIGO and Virgo, by detecting abundant gravitational waves coming from 10 to 30 solar-mass black hole binaries.

Another vital role of B-DECIGO is to verify the various technologies required for DECIGO. The technologies include the foreground clean-up [24]. DECIGO will detect many gravitational wave signals coming from neutron star and black hole binaries. We have to identify these signals and remove them from the observation data one by one to reach the primordial gravitational wave signals finally. This foreground clean-up technology is, in principle, possible, but it requires a demonstration using the real observation data. B-DECIGO will provide a framework for this demonstration.

## 4. Schedule of DECIGO and B-DECIGO

We plan to launch B-DECIGO as a precursor to DECIGO in the 2030s to demonstrate the technologies required for DECIGO, as well as to obtain fruitful scientific results to expand the multi-messenger astronomy further. Then we hope to launch DECIGO at a later time, incorporating lessons learned from B-DECIGO.

We are also considering the possibility of having two missions in advance to demonstrate the technologies required for B-DECIGO. The first mission is to demonstrate the FP cavity operation with two super-small spacecraft in space. The two spacecraft have two mirrors, each of which is attached to each spacecraft. The two mirrors form a FP cavity, which is illuminated by laser light. We will control the FP cavity on resonance by feeding the cavity error signals back to the thrusters of the spacecraft. We are thinking of doing this mission in collaboration with the spacecraft engineering group, formation-flight group, infrared-red astronomy group, and X-ray astronomy group. We hope to launch this mission around 2024.

The second one is to demonstrate the combination of the drag-free operation and the FP cavity operation with two or three small spacecraft in space. B-DECIGO requires technologies of the drag-free system and the FP cavity system. These two technologies are, in principle, compatible, but it is crucial to demonstrate this compatibility in the real mission. The formation-flight group is now considering leading this mission together with the DECIGO group and other groups. We hope to launch this mission in the late 2020s.

## 5. Conclusions

DECIGO aims at delivering a wide range of science. This includes the detection of the primordial gravitational waves to verify and characterize the inflation of the universe, measurement of the acceleration of the expansion of the universe to provide reliable information for dark energy, the prediction of the accurate time and direction for the electromagnetic follow-up observations, and other astrophysical and cosmological sciences. B-DECIGO is a scientific pathfinder mission of DECIGO. It aims at obtaining astrophysical sciences as well as demonstrating the technologies required for DECIGO. We hope to launch B-DECIGO in the 2030s. We are considering the possibility of having two small-size missions in the 2020s to demonstrate the technologies necessary for B-DECIGO. We believe that in the future B-DECIGO and then DECIGO will expand the gravitational wave astronomy and physics to full bloom.

## Acknowledgments

We would like to thank David H. Shoemaker for the English editing. This work was supported by Daiko Foundation, JSPS KAKENHI Grant Number JP19H01924.